\documentclass[aps,twocolumn,amsmath,amssymb,preprintnumbers]{revtex4}
\usepackage{amsmath} \usepackage{amsfonts} \usepackage{amssymb}
\usepackage{titlesec}
\usepackage{bbm}
\usepackage{epsfig}
\usepackage{graphics}
\usepackage{graphicx}
\newcommand{\be}{\begin{equation}}
\newcommand{\ee}{\end{equation}}
\newcommand{\ba}{\begin{eqnarray}}
\newcommand{\ea}{\end{eqnarray}}
\newcommand{\nn}{\nonumber}
\newcommand{\kr}{\rangle}
\newcommand{\kl}{\langle}
\newcommand{\m}{(x;\mu)}
\newcommand{\tr}{{\rm tr}}

\newcommand{\lin}[2][x]{(#1;#2)}
\newcommand{\gr}[2]{\mathrm{#1}(#2)}
\titleformat{\subsection}[block]{\normalfont\bfseries}{\thesubsection.}{1ex}{}
\titlespacing{\subsection}{0pt}{10pt}{1pt}[0pt]
\titleformat*{\section}{\large\bfseries}
\renewcommand{\thesubsection}{\arabic{subsection}}

\begin{document}

\title[ ]{Linear lattice gauge theory}

\author{C. Wetterich}
\affiliation{Institut  f\"ur Theoretische Physik\\
Universit\"at Heidelberg\\
Philosophenweg 16, D-69120 Heidelberg}

\begin{abstract}
Linear lattice gauge theory is based on link variables that are arbitrary complex or real $N\times N$ matrices, in distinction to the usual (non-linear) formulation with unitary or orthogonal matrices. For a large region in parameter space both formulations belong to the same universality class, such that the continuum limits of linear and non-linear lattice gauge theory are identical. We explore if the linear formulation can help to find a non-perturbative continuum limit formulated in terms of continuum fields. Linear lattice gauge theory exhibits excitations beyond the gauge fields. In the linear formulation the running gauge coupling corresponds to the flow of the minimum of a ``link potential''. This minimum occurs for a nonzero value of the link variable $l_0$ in the perturbative regime, while $l_0$ vanishes in the confinement regime. We discuss a flow equation for the scale dependent location of the minimum $l_0(k)$. 

\end{abstract}

\maketitle

\section{Introduction}
\label{Introduction}

The standard formulation of lattice gauge theories is based on link variables that are elements of the gauge group \cite{Wi}. 
For a gauge symmetry $\textrm{SU}(N)$ or $\operatorname{SO}(N)$ these are special
unitary or orthogonal matrices, respectively.
Due to the constraints of unitarity or orthogonality we may call such models ``non-linear lattice gauge theories'', in analogy to the non-linear $\sigma$ models.
In contrast, linear lattice gauge theories are based on arbitrary complex or real matrices for the link variables. 
They therefore contain additional degrees of freedom, similar to the ``radial
mode'' that is added if one changes from a non-linear $\textrm{O}(N)$-$\sigma$-model to a linear model with the same symmetry.

The additional degrees of freedom do not imply that linear lattice gauge theory differs from the standard non-linear version. Linear lattice gauge theory may be obtained from a usual lattice gauge theory by a reformulation on block lattices. For example, one may consider in two dimensions a block lattice where the block-links correspond to diagonals of the plaquettes of the original lattice. Denoting the unitary matrices on the four links around a given plaquette by $U_1,U_2,U_3,U_4$, a block link variable can be defined by
\be\label{I1}
L_{12}=\frac12(U_1U_2+U^\dagger_4U^\dagger_3).
\ee
It has the correct gauge transformation properties with unitary transformations at each end of the link. However, $L_{12}$ is not a unitarity matrix any more, $L_{12}L^\dagger_{12}=\frac14(2+U_1U_2U_3U_4+U^\dagger_4U^\dagger_3U^\dagger_2U^\dagger_1)$. (Only one of the diagonals is used for each plaquette, the block lattice distance is a factor $\sqrt{2}$ larger than the original lattice distance.) This simple property has an analogue in the Ising model where the block spins have no longer unit length. 

In this paper we do not aim to perform an explicit computation of block spin transformations. We rather take the property that block link variables are no longer unitary as a motivation for an exploration if linear lattice gauge theory could help to find a non-perturbative continuum limit for lattice gauge theories, formulated in terms of continuum fields. For a confining theory such a continuum limit necessarily involves fields different from the perturbative continuum gauge fields. The additional degrees of freedom in the linear formulation could be useful in order to account for such fields. This is precisely what happens in simpler theories as the Ising model or non-linear $\sigma$-models. The continuum limit is described by a linear $\varphi^4$-theory.

Linear lattice gauge theories have been explored since the early days of lattice gauge theories \cite{Pat, GM}. (The present paper shows overlap with this work in several aspects.) 
They have been left aside in the following, probably because their usefulness was not obvious as compared to the more economical non-linear formulation.
Recently, it has been proposed that lattice gauge theories can be based on scalars or fermions instead of link variables \cite{CWS}. 
Then the ``fundamental degrees of freedom'' are scalar or fermionic site variables, while gauge bosons
arise as collective or composite fields. 
The associated composite link variables are scalar or fermion bilinears which do not obey constraints, 
such that an investigation of linear lattice gauge theory becomes necessary.

A particular motivation for our investigation of linear lattice gauge theories arises from the close analogy 
of non-abelian gauge theories in four dimensions and non-abelian non-linear $\sigma$-models in two dimensions \cite{NSM1,NSM2,NSM3,NSM4,NSM5,NSM6}.
Both theories are asymptotically free and have a strong interaction regime where a mass scale is generated by dimensional transmutation. 
For the two dimensional non-linear $\sigma$-models the strong interaction
regime has found a simple description by means of functional renormalization. It corresponds to the linear $\sigma$-model without spontaneous symmetry breaking \cite{S2Da,CWFRG,S2Db,GW}.

Let us look at the two-dimensional scalar models in more detail. 
Within the linear description, the microscopic action for the non-linear model can be realized by a potential with a minimum located at a non-zero field value. 
The quartic scalar coupling is taken to infinity such that the potential effectively induces a constraint for the scalars 
$\phi_i (x)$, $\sum_i \phi_i^2(x)=\kappa_\Lambda, \: i=1,...N, \: N\geq3$. 
The remaining modes are the Goldstone bosons from the spontaneous breaking  of $\textrm{SO}(N)$ to $\textrm{SO}(N-1)$. 
They describe a non-linear $\sigma$-model with $\textrm{SO}(N)$ symmetry.
The coupling of the non-linear model is related to the location of the minimum in the linear model by $g^2=(2\kappa)^{-1}$. 
Functional renormalization can follow the flow of the potential minimum 
$\kappa(k)$ as a function of some infrared cutoff $k$, starting at some microscopic scale $k=\Lambda$ with $\kappa(\Lambda)=\kappa_\Lambda$. 
For the linear model in two dimensions the flow equation
for $\kappa$ matches precisely the well-known flow of $g(k)$ in the non-linear model, according to $g^2(k)=(2\kappa(k))^{-1}$. 
This holds for large $\kappa$ or small $g^2$ where perturbation theory is valid in the non-linear model.
Furthermore, one finds in the linear model that $\kappa(k)$ reaches zero for some particular scale $k_s>0$.
At this scale the non-abelian coupling formally diverges, $g(k\to k_s)\to \infty$.

In the linear model the flow of the potential can be followed also for $k<k_s$. The minimum of the potential remains at zero, $\kappa(k<k_s)=0$, 
such that the model shows no spontaneous symmetry breaking.
A mass term $m^2(k)$ is generated for the $N$ scalar modes -- it is equal for all modes according to the $\textrm{SO}(N)$ symmetry.
The running of $m^2(k)$ effectively stops once $k$ gets smaller than $m(k)$. The physical spectrum contains $N$ scalars with equal mass, $m\sim k_s$.

In this paper we explore the possibility that a similar mechanism can be found for the transition from weak couplings to the strong coupling regime of non-abelian gauge theories in four dimensions.
Linear lattice gauge theory indeed involves a potential for the link variables $L(x;\mu)$. Its minimum occurs for $L$ proportional to the unit matrix, $L(x;\mu)=l_0$.
One finds that the gauge coupling obeys $g^2=2/l_0^4$, in close analogy to the relation $g^2=(2\kappa)^{-1}$ for the two-dimensional $\sigma$-models. 
We compute the flow of the location of the minimum $l_0(k)$ by functional renormalization.
For large $l_0$ this reproduces indeed the one-loop running of the gauge coupling. According to this approximation $l_0(k)$ reaches zero at some value $k_s>0$, 
again similar to the two-dimensional $\sigma$-model. We therefore argue that the strong coupling or confinement regime of four-dimensional non-abelian gauge theories 
can be described by the continuum limit of linear lattice gauge theory in the symmetric phase, i.e. with $l_0(k<k_s)=0$.
Excitations around this minimum are massive and can be associated with glueballs.

The important new aspects of the present paper concern the relation between asymptotic freedom and confinement in linear lattice gauge theories. We establish a region in parameter space for which linear lattice gauge theories are in the same universality class as standard (non-linear) lattice gauge theories with small perturbative short-distance values of the gauge coupling. In particular, we present a limiting case where both types of models coincide. In view of the connection between the flow of the minimum of the link potential and the running gauge coupling we compute the dominant contribution to the flow of the link potential for large $l_0$. This permits us to relate the weak - and strong coupling regimes of linear lattice gauge theories quantitatively. Furthermore, our discussion of reduced symmetries $SU(N)$ instead of $SU(N)\times U(1)$ reveals interesting aspects of the role of the center symmetry $Z_N$. 

The main focus of the present paper is the region in parameter space for which linear lattice gauge theory is in the same universality class as standard lattice gauge theory. It therefore does not concern a ``new gauge theory'', but rather an investigation of the possible formulation of a continuum limit of standard gauge theories. Nevertheless, our approach offers additional perspectives. The parameter space of linear lattice gauge theories is sufficiently large in order to encompass universality classes that differ from the confining $SU(N)$ theories. For example, it can describe gauge theories with ``spontaneous gauge symmetry breaking. In the future, linear lattice gauge theory may also be used for an exploration of the boundaries of the standard universality class of confining gauge theories.

Our paper is organized as follows: A simple action for linear lattice gauge theories is presented in section \ref{Action for link variables}. 
It contains up to four powers of the link variables $L(x;\mu)$, with a link potential and a covariant derivative term. 
In section \ref{Unitarity link variables} we decompose an arbitrary complex link variable as a product of a hermitean matrix $S(x;\mu)$ and a unitary matrix $U(x;\mu)$. 
Gauge bosons are related to $U(x;\mu)$ in the usual way. The factor $S(x;\mu)$ contains additional fields and we concentrate on scalars $S(x)$. 
Both $S(x;\mu)$ and $S(x)$ transform homogeneously under gauge transformations.
We compute masses and kinetic terms for the singlet and adjoint scalars contained in $S(x)$.

In section \ref{Limit of standard non-linear lattice gauge theory} we discuss limiting values of the parameters characterizing the link potential $W_L(L)$.
In this limit the minimum of $W_L$ at $L=l_0$ is kept fixed, while the masses of all additional scalar fields tend to infinity.
The link potential acts then effectively as a constraint $L^\dagger(x;\mu)L(x;\mu)=|l_0|^2$.
Up to an overall normalization factor, the link variables become unitary matrices. One recovers the standard setting of non-linear lattice gauge theories.
In section \ref{SU(N) gauge symmetry and the standard model} we generalize the action in order to describe $\textrm{SU}(N)$-gauge theories or theories with gauge group 
$\textrm{SU}(N)\times\textrm{U}(1)$ with different values of the gauge couplings for the abelian and non-abelian factors.

We turn to the characteristic features of asymptotic freedom and confinement of non-abelian gauge theories in section \ref{Gauge fields glueballs and the confinement regime}.
We argue that for $l_0=0$ the gauge bosons are no longer propagating degrees of freedom, while the excitations $L(x;\mu)=l(x)$ can be associated with a massive glueball state.
In section \ref{Running minimum in linear gauge theory} we derive the flow equation for the scale dependent minimum $l_0(k)$ of the link potential.
For large $l_0$ this is dominated by gauge boson fluctuations. In lowest order, the flow of $l_0(k)$ describes the one-loop running of the gauge coupling.
Our conclusions are drawn in section \ref{Conclusions}.

\section{Action for link variables}
\label{Action for link variables}

We consider a hypercubic lattice in $d$-dimensions with lattice sites $x$ and lattice unit vectors $e_\mu ,\: \mu=1,...,d$. The lattice distance is denoted by $a, \: |e_\mu|=a$.
Links $(x;\mu)$ join the sites $x$ and $x+e_\mu$, and they have a direction (starting at $x$, ending at $x+e_\mu$). 
For each link we consider link variables  $L(x;\mu)$ that are complex or real $N\times N$ matrices, not subject to any constraint. 
With respect to local gauge transformations, the links transform as:
\be\label{A1}
L'(x;\mu)=V(x)L(x;\mu)V^\dagger(x+e_\mu).
\ee
For complex $L$ the matrices $V$ are unitary, $V^\dagger V=1$ and the gauge group is $\textrm{SU}(N)\times\textrm{U}(1)$.
For real $L$ one has orthogonal matrices ($V^\dagger=V^T,\:VV^T=1$), corresponding to an $\textrm{SO}(N)$ gauge symmetry.
These symmetries may be reduced, e.g. to $\textrm{SU}(N)$ instead of $\textrm{SU}(N)\times\textrm{U}(1)$, if the action is not invariant under the most general transformations \eqref{A1}.
The functional integral involves an integral over all link variables $L\m$,
\be\label{A2}
Z=\int\mathcal{D} L\exp\{-S_L\left[L \right]  \},
\ee
where appropriate source terms may be added.

The model is determined by the link action $S_L$.
This contains a ``link potential'' $W_L$ and a plaquette action $S_p$,
\be\label{17}
S_L=\sum_{{\rm links}}W_L\big(L(x;\mu)\big)+ S_p.
\ee
The ``link potential'' $W_L$ depends only on the matrix $L$ for one given link position $(x;\mu)$. We will use 
\ba\label{18}
W_L(L)&=&-\mu^2\rho+\frac{\lambda_1}{2}\rho^2+\frac{\lambda_2}{2}\tau_2,\nn\\
\rho&=&{\rm tr}(L^\dagger L)~,~\tau_2=\frac{N}{2}{\rm tr}(L^\dagger L-\frac1N\rho)^2,
\ea
where we assume $\lambda_1>0$ and $\lambda_2>0$ such that $S_L$ is bounded from below. 
(Higher order terms could be added, if necessary.)

The plaquette action
\be\label{2A}
S_p=\sum_{\mathrm{plaquettes}} \mathcal{L}_p(x;\mu,\nu)
\ee
plays the role of a kinetic term for the link variables.
Each term $\mathcal{L}_p(x;\mu,\nu)$ involves the four links of a plaquette $(x;\mu,\nu)$ with lattice points $(x,x+e_\mu,x+e_\mu+e_\nu,x+e_\nu)$ and the sum over plaquettes corresponds to 
$\sum_x\sum_\nu\sum_{\mu<\nu}$.
We take 
\ba\label{2B}
\mathcal{L}_p(x;\mu,\nu)&=& \frac{1}{4}\tr \big\{H^\dagger_{\mu\nu}(x)H_{\mu\nu}(x) \nn \\
                        &+& H^\dagger_{\nu-\mu}(x+e_\mu)H_{\nu-\mu}(x+e_\mu) \big\}
\ea
where summation over repeated indices is implied.

The quantities $H_{\mu\nu}$, $H_{\mu\,-\nu}$ are quadratic in the link variables,
\ba\label{2C}
H_{\mu\nu}(x)&=&L\m L(x+e_\mu;\nu) \\
&&-L(x;\nu)L(x+e_\nu;\mu),\nn
\ea
and 
\ba\label{2D}
H_{\nu\,-\mu}(x+e_\mu)&=&L(x+e_\mu;\nu) L^\dagger (x+e_\nu;\mu) \nn \\
&&-L^\dagger(x;\mu)L(x;\nu).
\ea
With respect to gauge transformations \eqref{A1} they transform as
\ba\label{2E}
H'_{\mu\nu}(x)&=&V(x)H_{\mu\nu}(x)V^\dagger(x+e_\mu+e_\nu),  \\
H'_{\nu\,-\mu}(x+e_\mu) &=& V(x+e_\mu)H_{\nu\,-\mu}(x+e_\mu)V^\dagger(x+e_\nu),\nn
\ea
such that $\mathcal{L}_p(x;\mu,\nu)$ is gauge invariant. We may define link variables with negative directions
\be\label{2F}
L(x+e_\mu;-\mu)=L^\dagger(x;\mu).
\ee
This makes it clear that $H_{\nu\,-\mu}(x+e_\mu)$ is obtained from $H_{\mu\nu}(x)$ by a $\pi/2$-rotation.  
The action \eqref{17} is therefore invariant under lattice translations and rotations.

The kinetic term $\mathcal{L}_p(x;\mu,\nu)$ contains two types of gauge invariants. The first invariant involves one link variable for each link around a plaquette,
\begin{align}\label{2G}
P\lin{\mu,\nu} =& P^*\lin{\nu,\mu} \\
=& \tr \left\{L\m L\lin[x+e_\mu]{\nu} L^\dagger\lin[x+e_\nu]{\mu}L^\dagger\lin{\nu} \right\}, \nn
\end{align}
while the second involves only two adjacent links with two variables for each link
\ba\label{2H}
Q\lin{\mu,\nu} &=& Q\lin{\nu,\mu} = Q^*\lin{\mu,\nu} \\
&=& \tr \left\{L\m L^{\dagger}\m L\lin{\nu}L^\dagger\lin{\nu} \right\}, \nn
\ea
In terms of these invariants one has
\begin{align}\label{2I}
\mathcal{L}_p\lin{\mu,\nu} =& -\frac{1}{2}\left[P\lin{\mu,\nu}+P\lin{\nu,\mu} \right] \\
+& \frac{1}{4}\big[Q\lin{\mu,\nu}+Q\lin[x+e_\mu]{\nu,-\mu}  \nn \\
+& Q\lin[x+e_\mu+e_\nu]{-\mu,-\nu}+Q\lin[x+e_\nu]{-\nu,\mu} \big]. \nn 
\end{align}

It is instructive to evaluate the action for a particular class of link configurations
\be\label{2J}
L\m=l\,U\m,
\ee
with $l$ a constant and $U\m$ unitary matrices, $U^\dagger U=1$.
One finds that the link potential $W_L$ and the invariant $Q$ are independent of $U\m$,
\ba\label{2K}
S_L &=& \sum_x\,dN \left(-\mu^2 l^2 + \frac{\lambda_1N}{2}l^4 \right) \\
&&+ \sum_{\mathrm{plaquettes}}l^4\left[N - \frac{1}{2} \left(P_U\lin{\mu,\nu}+P_U\lin{\nu,\mu} \right)\right],\nn
\ea
where $P_U$ obtains from $P$ by the replacement $L\to U$.
The second term corresponds to the Wilson action \cite{Wi} in the standard (non-linear) formulation of lattice gauge theories if we identify
\be \label{2L}
l^4 = \frac{\beta}{3}=\frac{2a^{d-4}}{g^2},
\ee
with $g$ the gauge coupling. 
This indicates that we will recover the universality class of standard lattice gauge theories if configurations of the type \eqref{2J} play a dominant role.

For positive $\lambda_1,\,\lambda_2$ the link potential is bounded from below.
Furthermore, $\mathcal{L}_p\lin{\mu,\nu}$ is positive semidefinite and the action is therefore bounded from below.
The minimum of the action occurs for constant link variables which realize a minimum of $W_L$.
For $\mu^2>0$ this ``ground state'' is simply given by all link variables proportional to the unit matrix
\be\label{2M}
L\m=l_0,
\ee
with
\be \label{2N}
\rho_0=N|l_0|^2=\frac{\mu^2}{\lambda_1}.
\ee
Without loss of generality we take $l_0$ to be real and positive. 
Of course, all configurations that can be obtained from the configuration \eqref{2M} by a gauge transformation \eqref{A1} are degenerate.

At this point the relation between linear and non-linear lattice gauge theories is similar to the relation between linear and non-linear $\sigma$-models \cite{JW}. 
The minimum of the potential in the linear formulation occurs for nonzero $l_0$ and the degrees of freedom of the non-linear model correspond to excitations around this minimum.
Furthermore, the linear model contains additional excitations beyond the ones described by the non-linear model.
In other words, the excitations around the minimum describe a standard lattice gauge theory with unitary link variables 
coupled to additional fields.

\section{Unitary link variables and ``link scalars''}
\label{Unitarity link variables}

In this section we establish the relation between linear lattice gauge theories and the usual lattice gauge theories based on unitary link variables. 
We first concentrate on complex link variables with gauge group $\textrm{SU}(N)\times\textrm{U}(1)$.
One can represent a complex $N\times N$ matrix $L$ as a product of a hermitean matrix $S$ and a unitary matrix (polar decomposition \cite{Dr}, see also ref. \cite{GM})
\be\label{19}
L(x;\mu)=S(x;\mu)U(x;\mu)~,~
S^\dagger=S~,~U^\dagger U=1.
\ee
The gauge transformation property
\ba\label{20}
S'(x;\mu)&=&V(x)S(x;\mu)V^\dagger(x),\nn\\
U'(x;\mu)&=&V(x)U(x;\mu)V^\dagger(x+e_\mu),
\ea
implies for $U(x;\mu)$ the same transformation property as for $L(x;\mu)$, while $S(x;\mu)$ involves only the gauge transformations at $x$. The fields
\ba\label{21}
S\m&=&l\m+A_S\m,\nn\\
l&=&\frac1N{\rm tr} S~,~{\rm tr} A_S=0~,~A_S^\dagger=A_S,
\ea
decompose into a singlet $l\m$ and an adjoint representation $A_S\m$. 
The singlet is invariant, while $A_S$ transforms homogeneously with respect to local gauge transformations at the point $x$. 

Without additional restrictions on $S$ the decomposition \eqref{19} is not unique. 
If two hermitean matrices $S_1$ and $S_2$ obey $S_2=S_1\tilde U$, with unitary $\tilde U$, we can equivalently use $S_1$ or $S_2$ in the polar decomposition, 
with suitably modified unitary matrices $U_1$ and $U_2$. 
In this case the local transformation $S\to S\tilde U,U\to\tilde UU$ leaves $L$ and therefore the action invariant. 
Expressed in terms of $S$ and $U$ the action will exhibit an additional gauge symmetry. 
This may be realized in a non-linear way since the existence of matrices $\tilde U$ typically depends on $S$, according to the condition $(S\tilde U)^\dagger=S\tilde U$. 
($\tilde U=-1$ is always a symmetry transformation.)

For each site $x$ we have $d$ fields $S\m$, one for each value of the index $\mu$. 
The precise properties of these fields with respect to the lattice symmetries are complicated.
For example, $\pi/2$-rotations can transform fields $S\m$ at different sites $x$ into each other. 
Suitable averages of fields over $\pi/2$-rotations can be associated with scalar fields, 
while the differences from these averages belong to other representations of the discrete lattice rotation group. 
Such differences between fields $S\m$ add substantial complication without involving qualitatively new aspects. 
We may neglect them here and concentrate on $S\m=S(x;\nu)=S(x)$, where $S(x)$ is associated with a scalar field. 
A more detailed discussion of the fields contained in $S\m$ can be found in the appendix.

The matrices $U\m$ play the role of unitary link variables which are familiar in lattice gauge theories. 
They are related to the gauge fields $A_\mu$ (represented here as hermitean $N\times N$-matrices) by
\be\label{28F1}
L(x;\mu)=S(x)U\m,~U\m=\exp\big\{ia A_\mu (x)\big\}.
\ee
Infinitesimal gauge transformations of $A_\mu$ involve the usual inhomogeneous term. 
Indeed, with $V(x)=\exp\big(i\alpha(x)\big)=1+i\alpha(x),\alpha^\dagger(x)=\alpha(x)$, eq. \eqref{20} implies (in lowest order in $a$)
\be\label{28F2}
\delta A_\mu=i[\alpha,A_\mu]-\partial_\mu\alpha.
\ee
Here we define lattice derivatives by 
\be\label{5A}
\partial_\mu f(x)=(f(x+e_\mu)-f(x))/a.
\ee

We next write the action in terms of the fields $U\m$ and $S(x)$.
With $LL^\dagger=SS^\dagger$ the link potential is independent of $U$, i.e. $W_L\big (L\m\big)=W_L\big(S\m\big)$. 
The unitary link variables appear only in the kinetic term ${\cal L}_p$ through the invariant $P$. For the action \eqref{17} this implies $S_L=S_g+S_W+S_A$, with 
\ba\label{22}
&&S_g=-\sum_{{\rm plaquettes}}
\big\{ l^2(x)l(x+e_\mu)l(x+e_\nu)Re\big(P_U(x;\mu,\nu)\big)\nn\\
&&\qquad -\frac N4\big[l^4(x)+l^2(x)l^2(x+e_\mu)+l^2(x)l^2(x+e_\nu)\nn\\
&&\qquad +l^2(x+e_\mu)l^2(x+e_\nu)\big]\big\}.
\ea
For $l(x)=l_0$ the ``gauge part'' of the action $S_g$ is precisely the plaquette action of standard lattice gauge theories \cite{Wi}
\ba\label{23}
S_g&=&-\frac{\beta}{3}\sum_{{\rm plaquettes}}\big \{{\rm Re}P_U(x;\mu,\nu)-N\big\}~,\nn\\
\ea
with gauge coupling given by equation \eqref{2L} for $l=l_0$. In particular, in four dimensions one has
\be\label{9A}
g^2=\frac{2}{l_0^4}.
\ee
In addition, $S_g$ contains derivative terms for the scalar singlet $l(x)$, which read in lowest order $a^2$
\be\label{SK1}
S^{(l)}_{\mathrm{kin}}=\frac12 N(d-1)a^2\sum_x\sum_\mu l^2(\partial_\mu l)^2+\dots
\ee

The potential part 
\be\label{37A}
S_W=d\sum_xW_L\big[l(x)+A_S(x)\big]
\ee
involves the scalar fields $l$ and $A_S$. Finally, the part $S_A$ contains covariant kinetic terms for the adjoint scalar $A_S$. 
It arises from $S_p$ and vanishes for $A_S=0$. This part can be found in the appendix. 

We conclude that for arbitrary complex $L$ and gauge group $\gr{SU}{N}\times \gr{U}{1}$ the action of linear lattice gauge theory 
describes gauge fields as well as scalars in the adjoint and singlet representations. 
Similarly, for real $L$ and gauge group $\gr{SO}{N}$ the matrices $S$ are symmetric and $U$ are orthogonal, $U^TU=1$. In this case $A_S$ corresponds to a traceless symmetric tensor representation.

\section{Limit of standard non-linear lattice gauge theory}
\label{Limit of standard non-linear lattice gauge theory}

We next show that the limit $\lambda_{1,2}\to\infty$ (fixed $l^2_0$) of linear lattice gauge theory results in the standard lattice gauge theory with unitary link variables.
For this purpose we choose parameters for the potential $W_L(S)$ for which a quadratic expansion around the minimum at $S=l_0$,
\ba\label{24}
W_L(S)&=&W_0+\frac12\bar m^2_l l^2_0(l-l_0)^2+\frac12\bar m^2_Al^2_0{\rm tr}(A^2_S)+\dots,\nn\\
\bar m^2_l&=&4N^2\lambda_1~,~\bar m^2_A=2N\lambda_2,
\ea
leads to large positive values $\bar m^2_l\gg 1,\bar m^2_A\gg1$. In order to extract normalized masses for the excitations we also need the kinetic terms for the scalars 
\ba\label{25}
S^{(l,A)}_{{\rm kin}}&=&\sum_x\frac12 Z_ll^2_0a^2\partial_\mu l(x)\partial_\mu l(x)\nn\\
&&+\frac12 Z_A l^2_0a^2{\rm tr}
\big\{\partial_\mu A_S(x)\partial_\mu A_S(x)\big\}.
\ea
Then the normalized mass terms read in the continuum limit 
\be\label{II}
m^2_l=\bar m^2_l/(Z_la^2),~m^2_A=\bar m^2_A/(Z_Aa^2).  
\ee
With eq. \eqref{SK1} one has $Z_l=N(d-1)$ and
\be\label{SK2}
m^2_l=\frac{4N\lambda_1}{(d-1)a^2}.
\ee
In the appendix we calculate also $Z_A$. In particular, for $d=4$ one finds for the scalar excitations
\be\label{14A}
m^2_l=\frac{4N\lambda_1}{3a^2},~m_A^2=\frac{2N\lambda_2}{3a^2}. 
\ee

For very large $m^2_l$ and $m^2_A$ the fluctuations of the scalar fields are strongly suppressed and give only minor corrections to the functional integral. 
In the limit $\bar m^2_l\to\infty$, $\bar m ^2_A\to\infty$ we approximate $S(x)$ by $l_0$. 
Then only $U\m$ remains as effective degree of freedom and we expect linear lattice gauge theory to give precisely 
the same results as non-linear lattice gauge theory for the corresponding value of $\beta=3l^4_0$. 
This extends to the more complicated structure of fields $S\m$. 

We conclude that our model of linear lattice gauge theory has a simple limit. 
For $\lambda_1\to \infty$, $\lambda_2\to\infty$, $\mu^2=N\lambda_1 l^2_0\to\infty$, with fixed $l^2_0$, the linear lattice gauge theory is equivalent to the standard (non-linear) 
lattice gauge theory with $\beta=3l^4_0$. Indeed, the limit $\lambda_1\to\infty,\mu^2=N\lambda_1 l^2_0$ can be interpreted as a constraint on the link variables
\be\label{30Aa}
\tr \{L^\dagger L\}=\tr S^2=N l^2_0.
\ee
For all values of S not obeying eq. \eqref{30Aa} the link potential diverges such that their contribution to the functional integral vanishes.
We can therefore replace $\rho$ by $NL_0^2$. Then the second limit $\lambda_2\to\infty$ leads to a second constraint
\be\label{30AB}
\tr \big\{(L^\dagger L)-l^2_0)^2\big\}=\tr\big\{ (S^2-l^2_0)^2\big\}=0. 
\ee
The solution of these two constraints reads $S^2=l^2_0$. 
This fixes $S$ to be of the form $S=l_0\tilde U,\tilde U^\dagger \tilde U=1,\:\tilde U^\dagger=\tilde U$. 
In turn, this implies the constraint that the link variables are unitary up to an overall constant, $L=l_0U$, such that we recover a standard $\gr{SU}{N}\times \gr{U}{1}$ lattice gauge theory. 

Starting from the limit $\lambda_{1,2}\to\infty$ we may lower the values of the couplings $\lambda_1$ and $\lambda_2$ while keeping $l^2_0$ fixed. 
For finite large values of $\bar m^2_l$ and $\bar m^2_A$ we still expect the model to be in the same universality class as standard lattice gauge theories. 
The long distance behavior will be characterized by the value of the renormalized gauge coupling. 
Its precise relation to the microscopic gauge couplings $g$ can typically be influenced by the presence of scalar fluctuations with masses of the order of the inverse lattice distance. 
Thus the relation \eqref{2L} can be modified for finite $\lambda_1,\lambda_2$, while the overall picture remains the same as long as these couplings are large enough.

\section{$\gr{SU}{N}$-gauge symmetry and the standard model}
\label{SU(N) gauge symmetry and the standard model}

So far we have shown that linear lattice gauge theories can realize $\mathrm{SU}(N)\times \mathrm{U}(1)$-gauge theories in the standard universality class. 
The argument for $\mathrm{SO}(N)$ gauge theories is completely analogous, using real instead of complex $N\times N$ matrices for the link variables.
For a realization of the $\mathrm{SU}(3)\times \mathrm{SU}(2)\times \mathrm{U}(1)$-gauge symmetry of the standard model 
we need a generalization to several $\mathrm{SU}(N)$-factors and different gauge couplings. 
We next show how to realize $\mathrm{SU}(N)$ gauge theories without the $\operatorname{U}(1)$ factor.

On the level of linear lattice gauge theory one may reduce the gauge group to $\mathrm{SU}(N)$ by explicit breaking of the $\mathrm{U}(1)$ symmetry. 
This can be achieved by the use of $\mathrm{SU}(N)$-invariants as
\be\label{42A}
d\m=\det \big(L\m\big).
\ee
According to eq. \eqref{A1} they transform under $\mathrm{U}(1)$ with a phase,
\be\label{42B}
d'\m=\exp \big\{i\big(\gamma(x)-\gamma(x+e_\mu)\big)\big\}d\m,
\ee
with 
\be\label{18A}
e^{i\gamma(x)}=\det V(x).
\ee
While $d\m$ is invariant under global $\gr{U}{1}$-transformations, the non-trivial phase under local transformations allows us to break the local $\gr{U}{1}$ symmetry
by adding to the link potential suitable terms containing $d\m$. Such terms will preserve the local $\mathrm{SU}(N)$ symmetry. 
For example, a term in the link potential of the type 
\be\label{42C}
W_d\sim -\big(d\m-d^*\m\big)^2
\ee
breaks the local $\gr{U}{1}$-gauge symmetry by favoring an alignment $d=d^*$. 
In addition, one may also have terms $\sim d+d^*$. 
(Note that global $\gr{U}{1}$ transformations are trivial for the models constructed here, since the link variables $L\m$ transform trivially as singlets.)

Let us add to the link potential $W_L$ a contribution 
\be\label{19A}
W_d = -\frac \nu2 (d+d^*)-\frac{\gamma}{4}(d+d^*)^2+\varepsilon dd^*,
\ee
with real parameters $\nu, \gamma$ and  $\varepsilon$.
Each factor $d$ involves $N$ links. Since the symmetry is reduced to $\gr{SU}{N}$ we decompose in eq. \eqref{28F1}
\ba\label{19B}
U\m &=& e^{i\varphi\m}\tilde U\m \nn \\
\det \tilde U\m &=& 1.
\ea
The special unitary matrices $\tilde U\m$ contain the gauge bosons of $\gr{SU}{N}$.
The potential $W_d$ remains independent of $\tilde U$ since
\be\label{19C}
d\m = \det S\m e^{iN\varphi \m}.
\ee
For nonvanishing $\nu \text{ or } \gamma$ it depends, however, on $\varphi\m$ which corresponds to the gauge boson of the abelian $\gr{U}{1}$-factor.
This excitation becomes massive. 

We may consider configurations with $S=l$. This yields for $W_d$ the expression
\be\label{19D}
W_d = -\nu l^N \cos(N\varphi) - \gamma l^{2N}\cos^2(N\varphi)+\varepsilon l^{2N}. 
\ee
For $\nu>0,\:\gamma>0$ and positive $l$ the minimum with respect to $\varphi$ occurs for
\be\label{19E}
\varphi_0 = \dfrac{2\pi n}{N},\quad n \in \mathbb{Z}.
\ee
At the minimum the second derivative of $W_d$ is positive
\be\label{19F}
\left. \dfrac{\partial^2 W_d}{\partial \varphi^2} \right|_{\varphi_0} = N^2\left( \nu l^N + 2\gamma l^{2N} \right),
\ee
such that $\varphi$ indeed describes a massive degree of freedom. The potential for $l$ gets modified by $W_d$ and one obtains for $\varphi=\varphi_0$
\be\label{19G}
W_L (l) = -N\mu^2 l^2 + \frac{\lambda_1 N^2}{2}l^4 - \nu l^N + (\varepsilon-\gamma)l^{2N},
\ee
such that the value $l_0$ for its minimum is shifted.
(We assume $\varepsilon \geq \gamma $ such that $W_L(l)$ remains bounded from below for arbitrary $N$.)
For the limit $\lambda_1 \to \infty$, $\lambda_2 \to \infty$, $\mu^2=N \lambda_1 l_0^2$ and finite $\nu,\,\gamma,\,\varepsilon$ this shift goes to zero.
We can then simply replace $l \to l_0$ in eqs. \eqref{19D}, \eqref{19F}.

The minimum of the link potential with respect to $\varphi$ has an $N$-fold degeneracy, cf. eq. \eqref{19E}.
This corresponds to the spontaneous breaking of a discrete $Z_N$-symmetry for $l_0\neq 0$.
Indeed, the action \eqref{17} is invariant under a global phase transformation of all link variables,
\be\label{19H}
L'\m = e^{i\alpha}L\m.
\ee
(This transformation may be obtained by a suitable combination of local transformations \eqref{A1}.)
The minimum of the link potential only fixes $|l_0|$, such that any particular choice of phase, i.e. $l_0$ real and positive, spontaneously breaks the global $\gr{U}{1}$-symmetry.
(The global $\gr{U}{1}$-symmetry \eqref{19H} should not be confounded with global $\gr{U}{1}$-transformations of the type \eqref{A1} which leave the link variables invariant.)
For non-vanishing $\nu$ or $\gamma$ the global $\gr{U}{1}$-symmetry is reduced to a $Z_N$-symmetry, with $\alpha=2\pi n/N$.
This explains the $N$-fold degeneracy of the minimum.

For the special case $\nu=\gamma=0$, $\varepsilon\neq 0$ the local $\gr{U}{1}$-symmetry remains intact.
Indeed the link potential remains independent of $\varphi\m$ and the gauge bosons of the abelian $\gr{U}{1}$ factor remain massless.
In the presence of the term $\sim \varepsilon$ the model remains $\gr{SU}{N}\times \gr{U}{1}$-symmetric.
Only $l_0$ is shifted and the masses of $l$ and $A_S$ obtain corrections.
We can also add kinetic terms involving $d$ that retain the local $\gr{SU}{N}\times \gr{U}{1}$-symmetry, as
\begin{equation}\label{19I}
\begin{split}
K_d &= \frac{Z_d}{4}\mkern-18mu \sum_{\;\;\scriptscriptstyle{\mathrm{plaquettes}}}\!\!\!\!\!\!\left\{\left|d\m d\lin[x+e_\mu]{\nu}-d\lin{\nu}d\lin[x+e_\nu]{\mu}\right|^2 \right. \\
&+ \left. \left|d(x+e_\mu;\nu)d^*(x+e_\nu;\mu)-d\lin{\nu}d^*\m \right|^2 \right\}.
\end{split}
\end{equation}
We will see in the next section that this contributes to the kinetic term of the $\gr{U}{1}$-gauge bosons.
As a consequence, the gauge couplings of the $\gr{SU}{N}$ and $\gr{U}{1}$ groups will be different, as required for the electroweak group of the standard model $\gr{SU}{2}\times \gr{U}{1}$.

The gauge symmetry $\gr{SU}{3}\times \gr{SU}{2}\times \gr{U}{1}$ of the standard model of particle physics can be realized by adding two independent pieces in the action.
For the first piece the variables are complex $3\times3$ matrices and the link potential contains a term \eqref{19A} with $\nu,\gamma \neq 0$.
For the second piece one uses for the link variables complex $2\times 2$ matrices, with $\nu,\gamma = 0$ in the link potential and adding a term $K_d$ according to eq. \eqref{19I}. 
One may also realize the standard model by a spontaneously broken grand unified symmetry as $\gr{SO}{10}$, that can be realized by choosing for the links real $10\times 10$ matrices.
Fermions or additional scalars as the Higgs doublet can be implemented in a gauge invariant way in complete analogy to the standard formulation.
The unitary link variables $U\m$ are simply replaced by the linear link variables $L\m$.
Problems with fermion doubling and the implementation of chiral fermions remain the same as in the standard setting.

\section{Gauge fields, glueballs and the confinement regime}
\label{Gauge fields glueballs and the confinement regime}

Having set the stage and formulated our models we now come to the main topic of the present work, namely the connection between asymptotic freedom and the confinement regime within linear lattice gauge theory. Indeed, linear lattice gauge theories may provide for a rather simple qualitative understanding of the confinement regime in non-abelian gauge theories.
This is connected to the observation that for $l_0\to 0$ the gauge coupling diverges. Already in the early discussion of linear lattice gauge theories or ``dielectric lattice gauge theories'' it has been argued \cite{GM} that confinement can be shown in such models. 

\subsection{Gauge fields}

This issue becomes clear if we express the plaquette part of the action of linear lattice gauge theory,
\be\label{42D}
S_p=\frac14\sum_{\rm plaquettes} \tr \{H^\dagger_{\mu\nu}H_{\mu\nu}+H^\dagger_{\nu-\mu}H_{\nu-\mu}\},
\ee
in terms of the gauge fields $A_\mu(x)$. Expanding
\be\label{21A}
L\m = l_0 \exp\left\{iaA_\mu(x) \right\}
\ee
up to $a^2$ one obtains
\ba\label{37F1}
H_{\mu\nu}&=&H_{\nu-\mu}=ia^2l^2_0F_{\mu\nu},\nn\\
F_{\mu\nu}&=&\partial_\mu A_\nu-\partial_\nu A_\mu+i[A_\mu,A_\nu].
\ea
Summing over the plaquettes this yields for the continuum limit of $S_p$
\be\label{37F2}
S_p=\int_x\frac14a^{4-d}l^4_0\tr\{F_{\mu\nu}F_{\mu\nu}\}. 
\ee
The microscopic lattice gauge coupling is therefore given by the inverse of the fourth power of the expectation value $l_0$
\be\label{26}
g^2=\frac{2 a^{d-4}}{l^4_0}.
\ee

In the presence of a term $K_d$ \eqref{19I} the abelian gauge field $\varphi\m=aA^{(1)}_\mu(x)$ receives an additional contribution to its kinetic term
\ba\label{23A}
K_d &=& \int_x \frac{1}{4}a^{4-d}Z_d N^2 l_0^{2N} F^{(1)}_{\mu\nu}F^{(1)}_{\mu\nu},\nn \\
F^{(1)}_{\mu\nu} &=& \partial_\mu A_\nu^{(1)} - \partial_\nu A_\mu^{(1)}.
\ea
The abelian gauge coupling $g^{(1)}$ therefore differs from the non-abelian gauge coupling,
\be\label{23B}
(g^{(1)})^2 = \dfrac{g^2}{1+Z_d N^2 l_0^{2N-4}}.
\ee
In the following, we will concentrate on $\gr{SU}{N}$-gauge theories by adding $W_d$ with $\nu, \gamma \neq 0$.

The gauge coupling \eqref{26} is small for large $l_0$ such that lattice perturbation theory can  be applied for not too large distance scales. 
On the other hand, for small $l_0$ one has a large $g^2$ and a strong coupling expansion becomes valid. 
For a simple non-abelian gauge group (as $\gr{SO}{2N}$ or $\gr{SU}{N}$) all correlation functions are expected to decay exponentially in this case, and no non-trivial long distance behavior is expected. 
(If the gauge group has an abelian factor (as $\gr{SU}{N}\times \gr{U}{1}$) non-trivial long distance behavior may be associated to a Coulomb type interaction in the abelian sector.)

\subsection{Asymptotic freedom and confinement}

The central idea for the connection between asymptotic freedom and confinement within linear (lattice) gauge theory is the observation that the link potential \eqref{18}, 
or its corresponding continuum version for scalar fields $S(x)$, depends on the renormalization scale $k$. 
The parameters $\mu^2,\lambda_1$ and $\lambda_2$ (and similar for $\nu, \gamma, \varepsilon$) will be replaced by running renormalized couplings. 
As a result, also the location of the minimum of the potential at $l_0(k)$ will depend on the renormalization scale.
The connection between asymptotic freedom and confinement is established if $l_0(k=\Lambda)$ is large microscopically for $\Lambda=\pi/a$,
while for $k\leq k_s$ one finds a vanishing expectation value, $l_0(k\leq k_s)=0$. 
Here $k_s$ is of the order of the ``confinement scale'' $\Lambda_{\rm QCD}$ which characterizes an appropriate momentum scale where the renormalized gauge coupling has grown very large. We associate confinement with the property $l_0(k=0)=0$.

For four-dimensional non-abelian gauge theories the situation is analogous to the relation between the linear and non-linear non-abelian $\sigma$-models in two dimensions \cite{S2Da, S2Db, GW}.
It is worthwhile to recall the properties of these models since important conclusions for four dimensional gauge theories can be drawn. 
The coupling of the non-linear $\sigma$-model is given by the inverse of the expectation value $\kl \varphi\kr$ in the linear model. 
More precisely, $\kappa=\kl\varphi^\dagger_R\kr\kl \varphi_R\kr$ is the value for which the potential $V(\varphi)$ in the linear model takes its minimum and one has the relation $g^2=(2\kappa)^{-1}$.
(We denote by $\varphi_R$ renormalized fields.)

Including  the effect of fluctuations the microscopic couplings are replaced by running renormalized couplings. 
Then the running of $g^2$ in the non-linear model can be associated to the running of $\kappa$ in the linear model. 
One can study the fluctuation induced  change of the effective potential $V(\varphi)$ in the linear model by use of functional renormalization \cite{S2Da, S2Db, GW}. 
For this purpose one introduces an effective infrared cutoff $k$ in order to define the effective average action $\Gamma_k$ which includes the quantum fluctuations with momenta larger than $k$. 
The scale dependence of $\Gamma_k$, and correspondingly of the effective  average potential $V_k(\varphi)$, is governed by an exact functional differential equation with one loop structure \cite{CWFRG}. 
For a non-abelian $\gr{SO}{N}$ symmetry and large $\kappa$ one obtains in leading order of a derivative expansion for the $k$-dependence of the minimum of $V_k(\varphi)$ the flow equation
\be\label{27}
k\partial_k\kappa=\frac{N-2}{4\pi}.
\ee
With $g^2=(2\kappa)^{-1}$ this reproduces precisely the one loop result for the running of $g$ in the non-linear $\sigma$-model.

Starting at some ultraviolet scale $\Lambda$ with $\kappa_\Lambda$ eq. \eqref{27} implies that $\kappa(k)$ vanishes for a scale 
\be\label{24A}
k_s=\Lambda\exp\left\{ -\frac{4\pi\kappa_{\Lambda}}{N-2}   \right\}. 
\ee
This feature persists beyond the leading order in the derivative expansion \cite{GW}. 
For $k<k_s$ the minimum of $V_k(\varphi)$ is at $\varphi=0$.
Once all quantum fluctuations are included for $k\to 0$ no spontaneous symmetry breaking is present, in accordance with the Mermin-Wagner theorem \cite{MW}. The strong coupling regime of the non-linear $\sigma$-model is simply described by the symmetric regime of the linear $\sigma$-model.

\subsection{Flowing minimum of the link potential}

We propose that a similar description for the confinement regime of four-dimensional Yang-Mills theories is possible within linear gauge theories. 
The most important features can already be seen in a minimal version of the continuum limit which includes besides the gauge fields only the simplest scalar field $l(x)$, 
with microscopic potential at the scale $\Lambda$
\be\label{46A}
W_{L,\Lambda}(l)=-\mu^2_\Lambda N l^2+\frac{\lambda_\Lambda}{2}l^4,
\ee
where $\lambda=\lambda_1N^2$. Our task will be the computation of the scale dependent effective action $\Gamma_k$, and in particular the scale dependent effective potential $W_{L,k}$ \cite{CWFRG}. 
One may employ a quartic polynomial expansion around the minimum of the potential,
\be\label{46B}
W_{L,k}=\frac{\lambda(k)}{2}\big(l^2-l^2_0(k)\big)^2,
\ee
where $\lambda(\Lambda)=\lambda_\Lambda$ and $l^2_0(\Lambda)=\mu^2_\Lambda N/\lambda_\Lambda=\mu^2_\Lambda/(\lambda_{1,\Lambda}N)$. 
(In this expansion we can include in $l_0(k)$ and $\lambda(k)$ also the contributions of $W_d$, cf. eq. \eqref{19D} for $\cos N\varphi =1$. We take here even $N$ for simplicity. 
For odd $N$ an expansion in $l-l_0$ is more appropriate since a cubic term $\sim \nu(l-l_0)^3$ will be present.)
The confinement region is reached if $l_0(k)$ hits zero. For $l_0(k)=0$ one uses $W_{L,k}=(\bar m^2_k/2)l^2+(\lambda(k)/2)l^4$, with positive $\bar{m}_k^2$. This applies to the flow for $k<k_s$. 

One possible way of computing the scale dependence of $l_0(k)$ could be the lattice renormalization group. This would use effectively a type of ``block lattice''. We have argued in the introduction that a linear formulation may be advantageous for the flow of the lattice action even if one starts with a standard lattice gauge theory. For large $l_0$ lattice perturbation theory could be applied. The lattice renormalization flow is best done by numerical calculations. In the present work we take a different road by investigating the flow in a continuum quantum field theory which includes fields for the most important degrees of freedom of linear lattice gauge theories. This allows the use of non-perturbative functional renormalization. We will find the most crucial features already in this continuum formulation. 

In the simplest version we investigate gauge fields coupled to a scalar singlet $l(x)$, with a characteristic $l$-dependence of the kinetic term for the gauge fields. One may choose a normalization for the singlet field $l$ such that the continuum limit for the effective average action becomes for all $k$
\be\label{28}
\Gamma_k=\int_x\frac18 l^4 F^z_{\mu\nu}F^z_{\mu\nu}+W_k(l)+\frac12 Z_k(l)\partial_\mu l\partial_\mu l +\dots,
\ee
where $F^z_{\mu\nu}$ is the non-abelian field strength for the gauge fields, $F_{\mu\nu}=F^z_{\mu\nu}\lambda_z/2,\tr(\lambda_z\lambda_{z'})=2\delta_{zz'}$. 
This amounts to a standard normalization of the coupling between $l$ and the gauge fields, while the kinetic term for $l$ can take a non-standard form. 
Indeed, $l$ is dimensionless and $Z_k(l)$ has dimension mass$^2$, with microscopic value $Z_\Lambda(l)=3Nl^2/a^2$. 
The dots in eq. \eqref{28} denote terms involving higher derivatives of the gauge and scalar fields or additional fields that we neglect here.
For a first approach one may further approximate $W_L$ by eq. \eqref{46B} and replace the function $Z_k(l)$ by a constant $Z(k)=Z_k\big (l_0(k)\big)$. 
The truncation of the effective action has then only three parameters $\lambda(k),l_0(k)$ and $Z(k)$.
The computation of the flow equations for $\lambda(k),l_0(k)$ and $Z(k)$ follows standard procedures of functional renormalization in the effective average action formalism. 
Both $l(x)$ and $A_\mu(x)$ are treated here as unconstrained fields. 
(Besides the local gauge symmetry the effective action \eqref{28} has a discrete symmetry $l\to-l$.) 

We present in the next section a functional renormalization group computation of the running of $l^2_0(k)$ in leading order.
We find that for large $l_0$ and large $\lambda$ the minimum $l^2_0(k)$ decreases logarithmically for decreasing $k$ according to 
\ba\label{XABC}
l^4_0(k)&=&l^4_0(\Lambda)-4\bar\beta \ln \frac\Lambda k,\nn\\
\bar\beta&=&\frac{11N}{48\pi^2}.
\ea
With $g^2=2/l^4_0$ (for $d=4)$ this reproduces the standard running of the gauge coupling in one loop order. 
According to \eqref{XABC}, $l_0(k)$ vanishes for $k_s$, 
\be\label{28A}
k_s = \Lambda \exp \left\{-\frac{l_0^4(\Lambda)}{4\bar\beta}\right\} = \Lambda \exp \left\{-\frac{1}{2\bar\beta g^2(\Lambda)} \right\}.
\ee
This corresponds to the one-loop ``confinement scale'', $\Lambda_{QCD} = k_s$.

\subsection{Glue balls}

The strong coupling regime corresponds to $l_0(k)$ approaching zero. If for $k=0$ the minimum of $W_k(l)$ occurs for $l=0$ the action \eqref{28} no longer describes propagating gluons.
The term $\sim l^4 F^z_{\mu\nu}F^z_{\mu\nu}$ becomes a derivative interaction involving four powers of the scalar field $l$ and two, three or four powers of the gauge fields $A_\mu$, 
while no standard kinetic term for $A_\mu$ is present anymore. 
On the other hand, for positive $Z_0=Z_{k\to 0}(l=0)$ and $\bar m^2_0=\partial^2 W_{k\to 0}/\partial l^2|_{l=0}$, one finds that $l$ describes a scalar $(0^{++})$ glueball with mass $m_G=(\bar m^2_0/Z_0)^{1/2}$. 
This would account for the lowest excitation of a confined non-abelian gauge theory. 
For a renormalized glueball field $l_R=Z^{1/2}_k(l_0)l$ the scalar kinetic term has a standard normalization and $l_R$ has dimension of mass. 
Gauge fields appear now in a term $\sim Z^{-2}_k(l_0)l^4_R F^z_{\mu\nu}F^z_{\mu\nu}$. 
These remarks generalize to the case where $Z_k(l)$ vanishes for $l\to 0$ provided that $Z^{-1}(l)\partial^2 W/\partial l^2$ takes a finite positive value for $k\to 0,l\to 0$. 

The simple mechanism of a vanishing ground state value for $l$ in eq. \eqref{28} is an interesting candidate for a description of confinement by properties of the effective action in the continuum. It shows analogies to ``dielectric confinement'' \cite{KS,GM}.
At the present stage this picture is rather rough and additional degrees of freedom may have to be included in the continuum limit.
In particular the ansatz \eqref{28} does not account for the observation that the $Z_N$-symmetry discussed in the preceding section is spontaneously broken for $l_0\neq 0$ and restored for $l_0=0$.
This could be improved by extending the discussion to a complex field $l(x)$ which incorporates the phase on which the $Z_N$-symmetry acts, with terms of the type of eq. \eqref{19A}.
For non-zero temperature this could make contact to Polyakov loops \cite{Pol}, and, including quarks and mesons, to the rather successful Polyakov-quark-meson model \cite{SPW}.

\section{Running minimum in linear gauge theory}
\label{Running minimum in linear gauge theory}

In this section we discuss briefly the flow equation for the parameter $l^2_0(k)$. 
First we show that the one loop running for the gauge coupling, $\partial g/\partial t=-\bar\beta g^3$, correlates to the flow of the minimum of $W_k$ at $l_0(k)$
\be\label{29}
k\partial_kl_0=\frac{\bar\beta}{l^3_0}.
\ee
This running obtains directly from gauge boson loops and is related to the normalization of the scalar field $l(x)$ according to eq. \eqref{28}. 
The propagators and vertices of the gauge bosons depend on $l$ in our setting. 
Gauge boson loops therefore contribute to the flow of the effective potential, both directly and indirectly through the renormalization of $l$. 
Not surprisingly, the running of $l_0(k)$ can therefore directly reflect the one-loop beta-function of standard non-abelian gauge theories.

The relation between the field $l(x)$ normalized according to eq. \eqref{28} and some ``microscopic field'' $\bar l(x)$ depends on the renormalization scale $k$. The flow of $\Gamma_k$ at fixed $l$ is therefore computed in two steps: one first computes the flow a fixed $\bar l$, and subsequently makes a $k$-dependent change of variables in order to extract the flow at fixed $l$.

For large $l$ the dominant contribution from gauge boson loops is related to this $k$-dependent rescaling of $l$. 
It can be extracted from earlier  work \cite{ReuW} in a straightforward way.
We first keep a fixed field $\bar l(x)$ which coincides with $l(x)$ at the microscopic scale. 
In terms of $\bar l$ we allow for a function $\tilde Z_F(\bar l)$ multiplying the gauge boson kinetic term, 
\be\label{54Aa}
{\cal L}_{F,k}=\frac14\tilde Z_{F,k}(\bar l)\bar l^4 F^z_{\mu\nu}F^z_{\mu\nu}.
\ee
At the microscopic scale $\Lambda=\pi/a$ it obeys $\tilde Z_{F,\Lambda}(\bar l)=1/2$. 
From ref. \cite{ReuW} we infer the flow equation for $\tilde Z_F(\bar l)\bar l^4$ which reads in our simple truncation
\be\label{54AB}
k\frac{\partial}{\partial k}\big(\tilde Z_F(\bar l)\bar l^4\big)=2\bar \beta 
\left(1-\frac{5N}{24\pi^2\tilde Z_F(\bar l)\bar l^4}\right)^{-1}.
\ee
This equation is supposed to be valid for large enough $\bar l^2$. 

For large $\bar l^2$ the leading term is simply
\be\label{54AC}
k\frac{\partial}{\partial k}\tilde Z_F(\bar l)=\frac{2\bar\beta}{\bar l^4},
\ee
with solution 
\be\label{54AD}
\tilde Z_F(\bar l)=\frac12-\frac{2\bar\beta}{\bar l^4}\ln \frac{\Lambda}{k}.
\ee
The rescaled field $l$ is related to $\bar l$ by
\be\label{54AE}
l^4=2\tilde Z_F(\bar l)\bar l^4.
\ee
It can be used as long as $\tilde Z_F(\bar l)$ remains positive, which is the case of interest for large enough $\bar l$.

The flow equation of the potential $W_L$ at fixed $l$ is related to the one at fixed $\bar l$ by
\be\label{54AF}
k\frac{\partial}{\partial k}W_L(l)=k\frac{\partial}{\partial k}W_L(\bar l)-
\left. \frac{\partial W_L}{\partial l}k\frac{\partial}{\partial k}l\right|_{\bar l},
\ee
with 
\be\label{54AG}
\left. k\frac{\partial}{\partial k}l\right|_{\bar l}=\frac{1}{2l^3}k
\frac{\partial}{\partial k}\tilde Z_F(\bar l)\bar l^4=\frac{\bar \beta}{l^3}.
\ee
For the truncation \eqref{46B} this yields
\be\label{54AH}
k\frac{\partial}{\partial k}W_L=k\frac{\partial}{\partial k}W_L(\bar l)-2\bar\beta\lambda 
\left(1-\frac{l^2_0(k)}{l^2}\right).
\ee
If we neglect for a moment the first term $k\partial W_L(\bar l)/\partial k$ we can infer the flow of the location of the minimum $l^2_0(k)$ from the extremum condition
$\partial W_{L,k}/\partial l^2_{|l^2_0}=0$, which is valid for all $k$ and implies
\be\label{54AI}
\left. k\frac{\partial}{\partial k}\frac{\partial W}{\partial l^2}\right|_{l_0}+ \left. \frac{\partial^2 W}{(\partial l^2)^2}\right|_{l_0}
k\frac{\partial}{\partial k}l^2_0=0.
\ee
One finds from eq. \eqref{54AH}
\be\label{54AJ}
k\frac{\partial}{\partial k}l^2_0=-\frac1\lambda 
\left.\left(\frac{\partial}{\partial l^2}k\frac{\partial}{\partial k}W_L\right)\right|_{l_0}=
\frac{2\bar\beta}{l^2_0}.
\ee
This coincides with eq. \eqref{29} and leads to the one loop running of the gauge coupling according to eq. \eqref{26}. The bracket on the r.h.s. of eq. \eqref{54AB} amounts to higher terms in an expansion in $g^2$ and actually accounts already for more than $90 \%$ of the two-loop beta-function for the gauge coupling \cite{ReuW}. From eq. \eqref{54AH} we can also infer the contribution to the flow $\lambda$,
\be\label{54AK}
k\frac{\partial}{\partial k}\lambda=\frac{\partial^2}{(\partial l^2)^2}k
\frac{\partial}{\partial k}W_{L_{|l_0}}=\frac{4\bar\beta\lambda}{l^4_0}.
\ee

We next establish that the flow equation \eqref{54AJ} is the leading contribution to the flow of $l_0$ for the range of large $l_0$. Discussing the size of the non-leading contributions will shed light on the role of these fluctuations in the range of small $l_0$ where they can no longer be neglected. 

The flow of $\tilde Z_F(\bar l)$ receives also contributions from loops containing scalars in inner lines. 
The effective action \eqref{28} contains cubic vertices $\sim l^3$ involving two gauge fields and one scalar, as well as higher vertices.
The cubic vertices contribute to the flow of the inverse gauge boson propagator and therefore to the flow of $\tilde Z_F$. 
For large $l$ the gauge boson propagator scales $\sim l^{-4}$, and the scalar propagator $\sim Z^{-1}_k(l)\sim l^{-2}$. 
For a massless scalar this contribution would be similar to the contribution of the gauge boson loops, but with a suppression $\sim k^2/(Z_kl^2)$. Furthermore, one has a suppression due to the effective scalar mass term $\bar m^2=2\lambda l^2$. 
For large $\lambda$ and $l^2$ the scalar contribution becomes small and may be neglected. 
In contrast, for small $l$ and $\lambda$ the scalar contributions may actually dominate the flow of $\tilde Z_F$ such that eq. \eqref{54AC} remains no longer valid. 

Gauge boson loops also contribute directly to the flow of $W_L$, e.g. by generating in eq. \eqref{54AH} a flow $(k\partial W_L/\partial k)(\bar l)$. 
As compared to the dominant contributions described by eqs. \eqref{54AJ}, \eqref{54AK} these effects are suppressed by $k^4/\lambda$. 
Furthermore, there are additional contributions to the flow of the effective potential from scalar loops. 
They are suppressed, however, for large $\lambda$ due to a large renormalized scalar mass term $m^2_R(k)=2\lambda(k)l^2_0(k)/Z(k)$. 
For $m^2_R\gg k^2$ the contribution of heavy particles is suppressed by ``threshold functions'' which involve powers of $k^2/m^2_R(k)$. 
We conclude that for large $l_0$ and $\lambda$ the dominant contribution to the flow of $l_0(k)$ is indeed given by eq. \eqref{29}. 
Only in the strong coupling regime other contributions become important.
Since this concerns only scales in the vicinity of $k_s$ the perturbative estimate of $k_s$ in eq. \eqref{28A} remains a valid approximation.

In principle, the computation of the flow of the effective action in the truncation of eq. \eqref{28} is a straightforward task. It will require a numerical solution of the flow equations, however. We postpone this to future work, since a reliable estimate also needs an assessment if the truncation \eqref{28} remains sufficient for capting the most important qualitative features, or if extensions like the use of a complex field $l$ are needed.

\section{Conclusions}
\label{Conclusions}

Linear lattice gauge theories describe gauge bosons coupled to additional degrees of freedom. These additional degrees of freedom need not to be ``fundamental''. As one possibility they may merely be a convenient parametrization of standard lattice gauge theories on the level of coarse grained lattices or ``block lattices''. Alternatively, they could show up as relevant differences to the standard formulation. Which one of these two possibilities is realized amounts to the question to which universality class a given region in the parameter space of linear lattice gauge theories belongs.

The properties of the additional degrees of freedom are largely determined by an effective potential, whose parameters flow with the renormalization scale $k$. On the microscopic level this potential is given by the ``link potential'' for the unconstrained matrices defined on the links of a lattice gauge theory. In particular, typical masses $m(k)$ of the additional degrees of freedom depend on $k$. 
As long as the scale dependent masses $m(k)$ are large compared to the renormalization scale $k$, they only lead to small corrections to the dynamics of the dominant gauge bosons.

In particular, this holds if at the microscopic cutoff scale $k=\Lambda=\pi/a$ the masses are large, $m(\Lambda)\gg\Lambda$. 
In this case we are guaranteed that linear lattice gauge theory belongs to the same universality class as standard lattice gauge theories.
We have discussed limiting values of the parameters $\lambda_{1,2}\to\infty,\,\mu^2\to\infty$ for which linear lattice gauge theories coincide with standard lattice gauge theories.

Even if the microscopic ratios $\Lambda/m(\Lambda)$ are small, fluctuation effects may induce a flow for which $k/m(k)$ becomes of the order unity. Rather than ``integrating out'' the additional degrees of freedom it may be advantageous to keep them explicitly. The corresponding additional fields may be helpful to formulate a continuum limit that can account for confinement. We have shown that a ``linear gauge theory'' could indeed lead to a comparatively simple description of the physics of confinement. 

The connection between asymptotic freedom for $k\to\infty$ and confinement for $k\to 0$ is provided by the renormalization flow of the effective potential $W_L$ and its minimum at $l_0(k)$. We have computed the flow of $l_0(k)$ in an approximation that is valid for large enough $l_0$. In this approximation we find that $l_0$ flows from large values to zero as the renormalization scale $k$ decreases. The flow is logarithmic and the scale $k_s$ where $l_0$ gets small is therefore exponentially small as compared to the lattice cutoff $\pi/a$. Our approximation reproduces one-loop perturbation theory. In this approximation $k_s$ can be associated to the confinement scale $\Lambda_{\rm QCD}$. 

In the lattice formulation the minimum of the microscopic action \eqref{17} corresponds to a non-vanishing value for the link variables proportional to the unit matrix
\be\label{41}
L_0\m = l_0.
\ee
The expectation value of $L\m$ vanishes, however, due to quantum fluctuations which induce $l_0(k<k_s)=0$,
\be\label{42}
\langle L\m \rangle = 0.
\ee
This is consistent with the general property that local gauge symmetries cannot be broken spontaneously. (Eq. \eqref{42} is invariant under the transformation \eqref{A1}.)
Furthermore, our results suggest that eq. \eqref{42} also holds in the presence of gauge fixing.
This is in close analogy to the non-abelian $\sigma$-models in two dimensions where the Mermin-Wagner theorem forbids spontaneous symmetry breaking in the infinite volume limit.
For four-dimensional gauge theories the running of the gauge coupling is directly related to the logarithmic flow \eqref{29} of $l_0(k)$ 
from the microscopic value \eqref{41} to the macroscopic vacuum expectation value, which eventually vanishes for $k=0$.

So far our picture of confinement is only qualitative. The approximations are too rough in order to extract quantitative predictions for the glueball spectrum or similar properties.
It may be worthwhile to invest effort into a more detailed functional renormalization group study for the flow in the regime of small and vanishing $l_0(k)$.
This may also shed light on the interesting physics at nonzero temperature. Improved accuracy may be achieved by extending the truncation \eqref{28} for the continuum limit. Using a complex field $l(x)$ could capture the interesting physics related to the center $Z_N$-symmetry. If one wants to resolve further glueball states beyond the scalar one, further fields would be needed. It is not clear, however, if they play an important quantitative role. 

Besides an investigation of the standard ``confining'' universality class of $SU(N)$-gauge theories the general setting of linear lattice gauge theory can cover a much wider range of physical situations. In dependence on the parameters of the link potential we expect a rich phase diagram of linear lattice gauge theory. Various symmetry breaking patterns can be realized. As an example, we consider $N=10$ with real matrices on the links and $SO(10)$-gauge symmetry. The field $S(x;\mu)$ belongs to the $54$-dimensional traceless symmetric tensor representation. Concentrating on the scalar degrees of freedom the link potential becomes a standard scalar potential. A detailed discussion of scalar potentials and associated symmetry breaking patterns in $SO(10)$ can be found in ref. \cite{CWSO11} (and references therein). Particularly interesting is a spontaneous breaking of $SO(10)$ to $SU(4)_c\times SU(2)_L\times SU(2)_R$ by $A_S=diag (a,a,a,a,a,a,b,b,b,b)$, $6a+4b=0$. This group contains the gauge symmetries of 
the standard model. For large $a$ and $l_0$ one expects the gauge bosons of $SU(4)\times SU(2)\times SU(2)$ in the perturbative regime, while all other excitations are heavy and decouple. An effective 
transition to the full $SO(10)$-gauge symmetry occurs when $a$ goes to zero. ``Spontaneous symmetry breaking'' by the Higgs mechanism can be described within the setting of linear lattice gauge theories. (Further spontaneous symmetry breaking to $SU(3)_c\times SU(2)_L\times U(1)_Y$ may be achieved by adding to the model scalar variables on the lattice site, belonging to the $126$-dimensional representation of $SO(10)$.

Last but not least we emphasize that lattice simulations of linear lattice gauge theories seem possible with a reasonable effort.
One may establish in this way for which range of parameters the model belongs to the same universality class as standard lattice gauge theories.
Furthermore, a suitable definition of the link potential $W_L$ for coarse grained lattices should permit one to investigate the flow of its minimum 
and compare with the results of the present paper.

\section*{Appendix: Continuum limit of linear lattice gauge theory}
\renewcommand{\theequation}{A.\arabic{equation}}
\setcounter{equation}{0}

In this appendix we discuss the continuum fields contained in the link variables $L\m$. This sheds additional light on the continuum limit of the action \eqref{17}. We employ a decomposition similar to ref. \cite{CWS},
\ba\label{LG1}
L\m&=&S(x)\big(1-aC_\mu(x)\big)U\m,\\
U\m&=&\exp\big\{ia A_\mu(x)\big\}, \nn \\
A^\dagger_\mu&=&A_\mu,\:C^\dagger_\mu=C_\mu,\:S^\dagger=S. \nn
\ea
This identifies in eq. \eqref{19} $S\m=S(x)\big(1-a C_\mu(x)\big)$. The discussion in sect. \ref{Unitarity link variables} neglects $C_\mu(x)$. 
We observe that the decomposition \eqref{LG1} shows redundancy since the total number $2N^2$ of real functions contained in the complex $N\times N$ matrix $L\m$
is expressed by $N^2$ functions $A_\mu,N^2$ functions $C_\mu$ plus $N^2$ functions $S$. 
The same function $S(x)$ is shared by all links $L\m$ (for all $\mu$ at given $x$).

We may define covariant lattice derivatives for $S$ by
\be\label{LG2}
D_\mu S(x)=\frac1a\big\{U\m S(x+e_\mu)U^\dagger\m-S(x)\big\},
\ee
and similar for $D_\mu C_\nu(x)$. They transform homogeneously,
\be\label{LG3}
\big (D_\mu S(x)\big)'=V(x)D_\mu S(x)V^\dagger(x).
\ee
In the continuum limit one has in lowest order
\be\label{LG4}
D_\mu S=\partial_\mu S+i[A_\mu,S].
\ee
Eq. \eqref{LG2} yields the useful identity
\be\label{LG5}
U\m S(x+e_\mu)=\big[S(x)+a D_\mu S(x)\big]U\m
\ee
and similar for $C_\nu(x),S(x)C_\nu(x)$ or $S\m$.

This identity can be employed for computing the product of two neighboring links in different directions
\ba\label{LG6}
G_{\mu\nu}(x)&=&L\m L(x+e_\mu;\nu)\nn\\
&=&S(1-a C_\mu)\big(S-aSC_\nu+aD_\mu S(1-aC_\nu)\nn\\
&&-a^2SD_\mu C_\nu\big)U\m U(x+e_\mu;\nu).
\ea
Accordingly, one finds (no summation over $\mu,\nu$ here)
\ba\label{LG7}
&&G_{\mu\nu}G^\dagger_{\mu\nu}=\\
&&\quad S(1-aC_\mu)\big[(S+aD_\mu S)(1-aC_\nu)-a^2 SD_\mu C_\nu\big]\nn\\
&&\quad \times\big[(1-aC_\nu)(S+aD_\mu S)-a^2D_\mu C_\nu S\big](1-a C_\mu)S,\nn
\ea
and 
\ba\label{LG8}
&&G_{\mu\nu}G^\dagger_{\nu\mu}=\\
&&\qquad \quad S(1-aC_\mu)\big[(S+aD_\mu S)(1-aC_\nu)-a^2SD_\mu C_\nu\big]\nn\\
&&\qquad \times U\m U(x+e_\mu;\nu) U^\dagger(x+e_\nu;\mu)U^\dagger\m\nn\\
&&\qquad \times\big[(1-aC_\mu)(S+aD_\nu S)-a^2D_\nu C_\mu S\big](1-aC_\nu)S. \nn
\ea
(All fields besides $L$ and $U$ are taken at $x$.) The traces of the two last expressions are manifestly gauge invariant. 

With $H_{\mu\nu}=G_{\mu\nu}-G_{\nu\mu}$ we obtain for the plaquette term \eqref{2B}
\ba\label{LG9}
\tr H^\dagger_{\mu\nu}H_{\mu\nu}&=&
\tr \{ G^\dagger_{\mu\nu} G_{\mu\nu}+G^\dagger_{\nu\mu}G_{\nu\mu}\nn\\
&&- G^\dagger_{\mu\nu}G_{\nu\mu}-G^\dagger_{\nu\mu}G_{\mu\nu}\}.
\ea
It contains covariant derivatives of the fields $S$ and $C_\mu$. In addition, one has in eq. \eqref{LG8} the gauge covariant factor 
\ba\label{LG10}
p_{\mu\nu}&=&U\m U(x+e_\mu;\nu) U^\dagger (x+e_\nu;\mu)U^\dagger(x;\nu)\nn\\
&=&1+ia^2 F_{\mu\nu}-\frac{a^4}{2} F_{\mu\nu} F_{\mu\nu}+\dots
\ea
with 
\be\label{LG11}
F_{\mu\nu}=\partial_\mu A_\nu-\partial_\nu A_\mu +i[A_\mu,A_\nu].
\ee
Here we have omitted terms that do not contribute to $\tr H^\dagger_{\mu\nu}H_{\mu\nu}$ in order $a^4$. 

Up to order $a^4$ one obtains for $C_\mu=0$
\ba\label{LG12}
\tr H^\dagger_{\mu\nu}H_{\mu\nu}&=&a^2\tr\big\{S^2(D_\mu SD_\mu S+D_\nu S D_\nu S\nn\\
&&-D_\mu S D_\nu S-D_\nu S D_\mu S)\big\}\nn\\
&&+a^4\tr \big\{S^4 F_{\mu\nu}F_{\mu\nu}+i (D_\nu SS^2D_\mu S\nn\\
&&-D_\mu SS^2 D_\nu S)F_{\mu\nu}\big\}.
\ea
With respect to $\pi/2$-rotations the terms $\sim D_\mu SD_\nu S$ are odd and therefore vanish if we add the $\pi/2$-rotated piece $\tr H^\dagger_{\nu-\mu}H_{\nu-\mu}$.
As a result, one obtains for the continuum limit with $C_\mu=0$ and $d=4$
\ba\label{LG13}
S_p&=&\int_x\tr\left\{\frac{3}{2a^2}S^2D_\mu SD_\mu S+\frac14 S^4 F_{\mu\nu} F_{\mu\nu}\right.\nn\\
&&+\frac i2 D_\nu SS^2D_\mu S F_{\mu\nu}\Big\}.
\ea
For $S=l$ proportional to the unit matrix the last term vanishes and one recovers eqs. \eqref{37F2}, \eqref{25} with $Z_l=3N$. 
In addition, eq. \eqref{LG13} specifies the derivative terms for the scalar in the adjoint representation $A_S$ (cf. eq. \eqref{21}), e.g. $Z_A=3$ in eq. \eqref{25}. 
For terms involving $A_S$ the last term in eq. \eqref{LG13} needs not to vanish.

The potential part of the action of linear lattice gauge theory
\be\label{LG14}
S_W=a^{-d}\int_x\sum_\mu W_L\m
\ee
obtains from eq. \eqref{18} with 
\ba\label{LG15}
\rho\m&=&\tr \big\{S^2(x)\big(1-aC_\mu(x)\big)^2\big\}\\
\tau_2\m&=&\frac N2\tr \big\{\big[S^2(x)\big(1-aC_\mu(x)\big)^2\big]^2\big\}-\frac12\rho^2\m.\nn
\ea
This seems to imply linear terms in $C_\mu$, of the type $\tr\{S^2\sum_\mu C_\mu\}$ or $\tr\{ S^4\sum_\mu C_\mu\}$. 
Such terms would violate the rotation symmetry in the continuum limit. 
However, we have not yet taken into account that $S$ and $C_\mu$ are not independent unconstrained fields. 
Taking these constraints into account leads effectively to the vanishing of the terms linear in $C_\mu$. 

We may define $S$ in terms of the link variables as 
\be\label{L1}
S(x)=\frac{1}{4d}\sum_\mu\big\{L\m+L(x;-\mu)+h.c.\big\}.
\ee
This combination transforms as a scalar under $\pi/2$-rotations around $x$ and reflections and obeys $S^\dagger=S$. 
Evaluating eq. \eqref{L1} for $x$-independent $S$ and $C_\mu$ and for $U=1$ yields the relation
\be\label{L2}
S=S-\frac{a}{2d}\sum_\mu\{S,C_\mu\}.
\ee
One concludes for the anticommutator between $S$ and $C_\mu$
\be\label{L3}
\sum_\mu\big\{S(x),C_\mu(x)\big\}=f_d(x),
\ee
where $f_d(x)$ vanishes for constant $S,C_\mu$ and $A_\mu=0$. 
Therefore $f_d(x)$ has to contain derivatives of $S$ or $C_\mu$, and gauge covariance implies that these must be covariant derivatives. 
Omitting the derivative term $f_d$ this yields $\tr\{\sum\limits_\mu S^PC_\mu\}=0$ for arbitrary powers $P$.
In consequence, the rotation-symmetry-violating terms in eq. \eqref{LG14} vanish.

We note that we have not specified the transformation properties of $C_\mu$ under $\pi/2$-rotations and reflections.
This issue is somewhat involved and not needed for the present purposes. It is not obvious that the fields $C_\mu$ play a crucial role in the continuum limit. One may therefore approximate the continuum limit by setting $C_\mu=0$ and keeping only the gauge bosons and the various scalar fields contained in $S(x)$. Replacing the covariant lattice derivatives by covariant derivatives the microscopic form of the continuum action can be extracted from the formulae of this appendix. The couplings of this continuum version will flow. For the standard universality class of confining gauge theories they are all expected to flow to partial fixed points, with the gauge coupling as the only remaining marginal parameter. 

\vspace{2.0cm}\noindent

\bibliography{linear_lattice_gauge_theory}

\end{document}